\documentclass[sigconf, nonacm]{acmart}

\usepackage{svg}
\usepackage{enumitem}
\usepackage{caption}
\usepackage{subcaption}
\usepackage{multirow}
\usepackage{framed}
\usepackage{tabularx}

\AtBeginDocument{%
  }

\setlist[description]{leftmargin=0pt,labelindent=0pt}

\begin{document}

\title{A Metascience Study of the Low-Code Scientific Field\texorpdfstring{\footnotemark[1]}{}}

\author{Mauro Dalle Lucca Tosi}
\orcid{0000-0002-0218-2413}
\affiliation{%
  \institution{Luxembourg Institute of Science and Technology}
  \city{Esch-sur-Alzette}
  \country{Luxembourg}
}
\email{mauro.dalle-lucca-tosi@list.lu}

\author{Javier Luis Cánovas Izquierdo}
\orcid{0000-0002-2326-1700}
\affiliation{%
  \institution{IN3 -- UOC}
  \city{Barcelona}
  \country{Spain}
}
\email{jcanovasi@uoc.edu}

\author{Jordi Cabot}
\orcid{0000-0003-2418-2489}
\affiliation{%
  \institution{Luxembourg Institute of Science and Technology}
  \institution{University of Luxembourg}
  \city{Esch-sur-Alzette}
  \country{Luxembourg}}
\email{jordi-cabot@list.lu}

\renewcommand{\shortauthors}{Tosi et al.}

\begin{abstract}
In the last years, model-related publications have been exploring the application of modeling techniques across various domains. 
Initially focused on UML and the Model-Driven Architecture approach, the literature has been evolving towards the usage of more general concepts such as Model-Driven Development or Model-Driven Engineering.
More recently, however, the term ``low-code'' has taken the modeling field by storm, largely due to its association with several highly popular development platforms. 
The research community is still discussing the differences and commonalities between this emerging term and previous modeling-related concepts, as well as the broader implications of low-code on the modeling field.
In this paper, we present a metascience study of Low-Code. 
Our study follows a two-fold approach: (1) to analyze the composition and growth (e.g., size, diversity, venues, and topics) of the emerging Low-Code community; and (2) to explore how these aspects differ from those of the ``classical'' model-driven community.
Ultimately, we hope to trigger a discussion on the current state and potential future trajectory of the low-code community, as well as the opportunities for collaboration and synergies between the low-code and modeling communities.
\end{abstract}

%% jcanovasi %% I gave it a try but it was really hard to find the right terms :)



\maketitle

\renewcommand{\thefootnote}{\fnsymbol{footnote}}
\footnotetext[1]{This paper was originally published in the \textit{Journal of Object Technology}, vol. 24, no. 2, 2025. \url{http://dx.doi.org/10.5381/jot.2025.24.2.a10}}
\renewcommand{\thefootnote}{\arabic{footnote}} % Switch back to normal numbering

\section{Introduction}
\label{sec-intro}

In recent years, the ``Low-Code'' software development approach has gained significant traction. 
Low-Code promotes the development of software in a simplified manner, commonly involving a graphical user interface to specify the application and generators to create such application while requiring minimal coding skills from the user~\cite{waszkowski2019low}. 
While the term ``Low-Code'' was coined by Forrester Research in 2014~\cite{richardson2014new}, it first appeared in the scientific literature in 2017~\cite{baldwin2017hur}, and gained broader attention in the research community following  Zolotas \textit{et al.}~\cite{zolotas2018restsec} in 2018.

Despite its fast growth, the structure and characteristics of the Low-Code research community remain unexplored. 
Existing studies have investigated the novelty of Low-Code~\cite{bock2021search}, its research streams~\cite{naqvi2024low}, and its usability~\cite{pinho2023usability}. 
However, there are no comprehensive analyses of the Low-Code community itself.  
Such an investigation could benefit both new researchers seeking to enter the Low-Code field and experienced authors aiming to contribute to its growth or integrate Low-Code tools into their workflows to enhance software development efficiency.  

Long before Low-Code's emergence, the Software Engineering (SE) community has demonstrated the importance and benefits of adopting (standard) modeling paradigms to optimize the software development and maintenance processes~\cite{brambilla2017model, dzidek2008realistic}. 
Over the years, paradigms such as Model-Driven Architecture (MDA)~\cite{soley2000model}, Model-Driven Development (MDD)~\cite{selic2003pragmatics}, Model-Driven Engineering (MDE)~\cite{kent2002model}, Model-Based Architecture (MBA)~\cite{boehm1999conceptual}, Model-Driven Software Engineering (MDSE)~\cite{brambilla2017model}, and Model-Based Engineering (MBE)~\cite{brambilla2017model} have established themselves as de facto standards in the SE community. 
Based on their broad adoption and wide recognition by researchers and practitioners, we refer to them in this paper as traditional or classical Modeling paradigms.

Unlike traditional Modeling paradigms, Low-Code offers a much lower entry barrier and learning curve. 
Despite the potential technical differences between the two concepts, both Low-Code and traditional Modeling share commonalities and the goal to accelerate the creation of software systems~\cite{di2022low, cabot2020positioning, cabot2024thelowcode}. 
In fact, Low-Code is centered around Conceptual Modeling \cite{bock2021search} as are the traditional Modeling paradigms. 
Thus, enabling it to leverage decades of foundational research in the area.

Given these similarities, one might expect synergies between the two fields. 
Such synergies could enrich the Low-Code community with theoretical and practical insights from traditional Modeling, while revitalizing interest in Modeling through the growing attention and industry adoption of Low-Code. 
Understanding the connections and differences between these paradigms is crucial to identifying potential collaboration and growth opportunities for both communities.

Studying the evolution of a research field and its connections or divergences from related fields is a complex task.
The first step is to determine the appropriate analytical approach to track a field’s development and compare it with others.
One common method involves analyzing the research streams within a scientific field, identifying key streams \cite{naqvi2024low, tosi2021scikgraph}, and tracing their evolution over time \cite{dalle2022understanding}.
However, comparing these evolutions across different fields is challenging due to the difficulty of automatically identifying comparable research domains and the potential for bias in manual identification.
Alternatively, a more pragmatic approach relies on metadata from academic publications and artifacts to enable the quantitative comparison of different fields over time, offering a more scalable and systematic means of analysis.

In this sense, this paper presents a metascience study over the Low-Code field. 
In particular, (1) we examine the composition of the emerging Low-Code community, analyzing its size, diversity, preferred venues, among other characteristics; and (2) we investigate how this community differs from the traditional Modeling community in terms of people, venues, types of publications, and developed platforms. 
To this end, we analyze the evolution of the number of Low-Code and traditional Modeling publications, the correlation between the most productive and influential authors in both fields, the most popular venues and publication types for Low-Code articles, among other factors.

The paper is structured as follows.
Section~\ref{sec-methods} describes the methodology of our study, including the research questions, data collection, and threats to validity.
Section~\ref{sec-results} presents the answers to the research questions, and Section~\ref{sec-discussion} discusses some insights and recommendations derived from the data analysis.
Finally, Section~\ref{sec-conclusion} concludes the paper and outlines the future work.

\section{Methodology}
\label{sec-methods}
This section outlines the setup of our study. We begin by presenting the research questions, followed by the data collection and analysis processes employed to address them.

\subsection{Research Questions}
The primary objectives of our study are to: (1) characterize and analyze the composition and growth of the Low-Code research community, and (2) compare it with the traditional Modeling research community to investigate the potential impact of Low-Code in the overall Modeling community. 
To address these objectives, we aim to answer the following research questions:

\begin{description}[leftmargin=1em]
    \item[RQ1] What is the composition and growth of the Low-Code research field?
    To address this research question, we further subdivide it into the following:
    \begin{description}[leftmargin=1em]
        \setlength{\itemsep}{0.2em}
        \item[RQ1.1] How has the number of publications related to Low-Code evolved over time? 
        This research question aims to identify the starting point of Low-Code publications and their evolution over time. 
    
        \item[RQ1.2] What are the most common publication types and venues for Low-Code research? 
        This research question seeks to characterize the current typical target audience for Low-Code publications.
    
        \item[RQ1.3] How do Low-code publications contribute to the field of low-code?
        This question explores how Low-Code-related publications advance, investigate, or apply the field's concepts.
        
        \item[RQ1.4] What is the typical author profile in Low-Code publications?
        By analyzing co-authorship networks, number of publications, and h-index within the Low-Code field, this research question provides insights into the structure and sparsity of the Low-Code research community.
    
        \item[RQ1.5] How many Low-Code platforms and tools are available as open-source repositories, and which are the most popular? 
        This research question evaluates the popularity and adoption of Low-Code platforms among developers.
    \end{description}
    
    \item[RQ2] What is the impact of Low-Code on the traditional Modeling research community? 
    To address this research question, we propose the following sub-questions:
    \begin{description}[leftmargin=1em]
        \setlength{\itemsep}{0.2em}
        \item[RQ2.1] Are Low-Code publications affecting the number of traditional Modeling publications?
        To explore this question, we analyze the trends in the number of traditional Modeling publications and compare them with those in the Low-Code domain. 
        
        \item[RQ2.2] How many Modeling conferences and workshops aim to explicitly attract Low-Code publications? 
        This question seeks to understand how the Modeling community is engaging with Low-Code. 

        \item[RQ2.3] Are Low-Code publications present in Modeling conferences and workshops targeting them? 
        This sub-question aims to assess if Low-Code is indeed present within Modeling-focused venues. 
        
        \item[RQ2.4] How are authors distributed across Low-Code and traditional Modeling publications? 
        Here, we examine the profiles of authors involved in both paradigms by analyzing metrics such as publication counts and h-index.
        In particular, we compare relevant authors in each field and those contributing to both.  
    
        \item[RQ2.5] How many Low-Code open-source platforms and tools explicitly relate to traditional Modeling techniques? 
        This sub-question provides insights into developers' perceptions of incorporating traditional Modeling techniques into Low-Code development.
    \end{description}
\end{description}

To address each research question, we propose a metascience study involving the collection and analysis of publication histories from the Low-Code and Modeling paradigms.
The following sections describe the processes of data collection, cleaning, analysis, and visualization used to address our research questions.

\subsection{RQ1: Composition and growth of the Low-Code field}
\label{sec-methods:rq1}

\noindent \textbf{Data Collection}.
% To achieve the first main objective of this study, we started by addressing RQ1.1--RQ1.4.
To address RQ1.1--RQ1.4, we first collected a list of all publications related to Low-Code.
For this purpose, we relied on \textsc{Lens.org}\footnote{\url{https://www.lens.org/}}, a web platform that aggregates and provides publicly available scholarly metadata from diverse sources such as \textsc{ORCID}\footnote{\url{https://orcid.org/}}, \textsc{CrossRef}\footnote{\url{https://www.crossref.org/}}, and \textsc{PubMed}\footnote{\url{https://pubmed.ncbi.nlm.nih.gov/}}.
We built and executed the search query shown in Figure~\ref{fig:searchQuery} substituting ``TERM'' with our field of interest. 
As a result, we obtained 1,295 results when substituting ``TERM'' with ``low-code'' and 14 by ``\mbox{lowcode}''.
Note that the character ``-'' was automatically converted to a space during query execution and that the results were obtained after disabling the automatic stemming of the search terms. 
This step was performed on October 21$^\text{st}$, 2024, and the results reflect the publication history up to that date. 

\begin{figure}[htb]
    \[
    \begin{array}{c}
    \textrm{(title:("TERM") OR abstract:("TERM") OR} \\
    \textrm{keyword:("TERM") OR field\_of\_study:("TERM"))}
    \end{array}
    \]
    \caption{Search query used for data collection.}
    \label{fig:searchQuery}
    \vspace{-1em}
\end{figure}

To answer RQ1.5, we used the \textsc{GitHub API}\footnote{\url{https://docs.github.com/en/rest/quickstart?apiVersion=2022-11-28}} to identify repositories that declare themselves as Low-Code projects and aim to generate components of a software application, such as AI components, dashboards, or complete applications. 
These repositories were identified as the ones containing ``Low-Code'' as a keyword.
We further filtered the list to include only repositories with at least 50 stars and that were active --- defined as having their last commit within one year of the search date (November 17$^\text{th}$, 2024). Using this approach, we identified 301 repositories.

\medskip
\noindent \textbf{Data Cleaning \& Validation}.
In this step, we began by cleaning the data to answer RQ1.1--RQ1.4. 
First, we ensured that all documents in the collection of Low-Code publications were related to the Low-Code field and that their metadata was validated.
First, we merged the publications collected under the terms ``low-code'' and ``lowcode''. 
Next, we manually analyzed each publication to confirm it was related to the topic covered in this meta-analysis. 
This cleaning step was crucial, as it removed 397 publications that mentioned the searched term in unrelated contexts, e.g., ``This paper presents new methods for low code frequency and high code frequency testing.''~\cite{serra2005combined}. 
We then excluded publications that were unavailable (i.e., links with no accessible content), preprints that had already been published elsewhere (i.e., duplicates), and retracted articles. 
This second cleaning step removed 66 publication records, resulting in a curated Low-Code collection of 844 papers. 

During the validation of the metadata, we reviewed publication types to ensure the following:
(1) conference proceedings and journals published as books were not classified as book chapters;
(2) symposium, workshop, and conference papers were classified under ``conference proceedings''; 
and (3) bachelor's, master's, and doctoral theses were categorized as ``dissertation''. 
Additionally, we added missing author information to four publications and standardized publication venue names by removing edition and year details.

For the list of Low-Code repositories used to address RQ1.5, we cleaned the data by removing repositories that: (1) were not in English; (2) were used solely to host the source code of a published article; and (3) only listed Low-Code tools and platforms without contributing code.
This cleaning process removed 150 repositories from our initial list.

\medskip
\noindent \textbf{Data Analysis}. 
To address RQ1.1, we analyzed the number of Low-Code papers published annually until 2023.
We did not include scientific documents published in 2024 because the data collection process occurred before all Low-Code publications from 2024 had been indexed.

To answer RQ1.2, we counted and compared the number of Low-Code books, book chapters, journal articles, conference proceedings, preprints, dissertations, and other types of publications. 
Additionally, we counted and compared the number of Low-Code papers published by each journal and conference.

To answer RQ1.3, two authors manually categorized the Low-Code publications into one of the following groups: (1) \textit{Low-Code users}, for papers using Low-Code in their solutions; (2) \textit{Low-Code solutions}, for papers proposing frameworks, techniques, or solutions to improve the Low-Code area; (3) \textit{Low-Code platforms}, for papers introducing new platforms for Low-Code users; (4) \textit{Low-Code evolution}, for papers analyzing existing Low-Code publications; (5) \textit{Low-Code learning/teaching}, for papers focused on teaching or using Low-Code in education; and (6) \textit{Others}, for papers not fitting into these categories.

The manual categorization process was conducted as follows: one author annotated 50 randomly selected Low-Code articles based on their titles and abstracts to define the categories. 
A second author then re-annotated the same 50 articles without access to the initial annotations. 
By measuring the inter-rater reliability via the Cohen's kappa coefficient~\cite{cohen} we observed a ``Moderate agreement'' between the two annotators ($\kappa = 0.47$).
To further improve agreement, the authors discussed discrepancies in the initial set and proceeded to annotate the remaining 794 articles, setting aside items where disagreements were anticipated. 
After resolving disagreements and excluding 81 articles unavailable in English or other known languages by the authors, a total of 763 articles were successfully categorized.

For RQ1.4, we counted the number of Low-Code papers published by each author and calculated the authors' influence within the Low-Code field using their \textit{h-index within field}.
This metric is defined as the number of publications in the field where the author was cited at least as many times.
Additionally, we constructed a co-authorship graph with nodes representing authors and edges representing co-authorship. 
The nodes were positioned using the Fruchterman-Reingold force-directed algorithm~\cite{fruchterman1991graph} and were weighted, sized, and colored based on the number of articles each author had published.
Edges were weighted, sized, and colored based on the number of co-authored publications between two authors.

Finally, to answer RQ1.5, we counted the number of Low-Code repositories obtained after data cleaning.
Furthermore, we developed a Streamlit App~\cite{richards2023streamlit} to automate the data collection process described earlier in this section and made it publicly available\footnote{\url{https://oss-lowcode-tools.streamlit.app/}}.

\subsection{RQ2: Impact of Low-Code in the Modeling community}

\noindent \textbf{Data Collection}.
To gather the data for answering RQ2.1 and RQ2.4, we collected a list of publications related to the following traditional Modeling paradigms: Model-Driven Architecture (MDA), Model-Driven Development (MDD), Model-Driven Engineering (MDE), Model-Based Architecture (MBA), Model-Driven Software Engineering (MDSE), and Model-Based Engineering (MBE). 
Similar to the data collection process for RQ1, we obtained our list of publications by running the query shown in Figure~\ref{fig:searchQuery} in \textsc{Lens.org}. 
This process resulted in the following publication counts: 3,205 results for ``model-driven architecture'', 3,985 for ``model-driven development'', 5,196 for ``model-driven engineering'', 200 for ``model-based architecture'', 435 for ``model-driven software engineering, and 1,084 for ``model-based engineering''.

To collect the data for answering RQ2.2, we began by compiling a list of conferences related to traditional Modeling that either took place or were scheduled for 2024. 
This list included all events (conferences, workshops, symposiums, etc.) retrieved from the ICORE Conference Portal~\cite{icore} and the \textsc{Research.org} website that contained the term ``Model'' in their name. 
Additionally, we included events listed on the Domain-Specific Modeling (DSM) forum\footnote{\url{http://www.dsmforum.org/events.html}}. 
We also manually supplemented the list with popular Software Engineering conferences and their co-located events that are known to cover traditional Modeling topics but were not listed as 2024 DSMs event and did not include ``Model'' in their name. 
Examples of such conferences include FSE, ICSE, and CAiSE. 

After filtering out unrelated events, we compiled a final list of 64 conferences and 84 workshops. 
From this list of venues, we manually extracted relevant text fragments from their websites, focusing on sections describing their calls for papers, relevant topics, event descriptions, tutorials, and additional calls for contributions.

To address RQ2.3, we downloaded the abstracts of papers published in the ECMFA and MODELS conferences, along with those from companion events of MODELS (MODELS-C).
These venues were chosen based on the results from RQ2.2 and RQ1.2, which identified ECMFA and MODELS as two of the few Modeling conferences with Low-Code as a primary topic of interest (cf. Section~\ref{sec-results-rq22}), and MODELS-C as the set of events with the highest number of Low-Code publications (cf. Section \ref{sec-results-rq12}).    
We first extracted the list of papers published in these target venues in JSON format from the DBLP~\cite{dblp} website.
Next, we wrote a Python script to retrieve their DOI information. 
Using the extracted DOIs, we created a \textsc{Lens.org} collection of papers.
Finally, we downloaded the collection, which included the articles' abstracts, for use in the subsequent data analysis.

For RQ2.5, we used the same data collected to answer RQ1.5.

\medskip
\noindent \textbf{Data Cleaning \& Validation}.
For RQ2.1 and RQ2.4, we did not perform the same data cleaning and validation processes as for RQ1.1 and RQ1.4, given the lower likelihood of finding the searched terms in unrelated contexts and the impracticality of manually cleaning such larger collections.

Regarding the list of conferences gathered for RQ2.2, we identified and excluded eight conferences that contained the term ``Model'' in their name but were unrelated to the studied topic (e.g., ``Computational Models of Argument'').

For RQ2.3, no data cleaning or validation process was deemed necessary. 

Finally, for RQ2.5, we further cleaned the list of Low-Code repositories identified when answering RQ1.5 by removing those that did not contain the term ``model'' or its derivatives. Thereby, selecting only Low-Code tools and platforms explicitly related to traditional Modeling.

\medskip
\noindent \textbf{Data Analysis}. 
To address RQ2.1, we analyzed the number of traditional Modeling papers published annually. We then compared the trends of Low-Code publications (from RQ1.1) with those from individual Modeling paradigms and with all paradigms combined.

For RQ2.2, we identified the number of Modeling conferences and co-located events that actively target the Low-Code community. 
Specifically, we searched for venues mentioning the terms ``low-code'', ``lowcode'' or ``low code'' on  their websites. 
We then examined where on the website those terms were mentioned, which provided insights into the types of Low-Code contributions the venue was expecting.
Finally, by analyzing the context in which the Low-Code terms appeared, we determined whether Low-Code was a primary topic of interest for the venue or simply one of many keywords used to describe general interests.

To answer RQ2.3, we focused on identifying the presence of the Low-Code topic in Modeling venues that actively encourage Low-Code-related submissions.
As a proxy, we analyzed the relevance of Low-Code-related keyphrases in the abstracts of papers presented at the target conferences.
We aggregated the abstracts of papers by venue (ECMFA, MODELS, or MODELS-C) and year of publication into separate documents. 
We then extracted the keyphrases for each of those documents using Topic-Rank \cite{bougouin2013topicrank}, C-Rank \cite{tosi2019c}, and WordCloud \cite{mueller_2024_14062883}. Based on our observations, WordCloud produced the best results and was therefore used for the remainder of this paper. 

By analyzing the extracted keyphrases, we aimed to determine the relevance of the Low-Code topic. This involved comparing the weights of keyphrases such as ``low code'', ``low-code'' ``low'', and ``code'' to other relevant keyphrases within the same conference and year. 

For RQ2.4, we counted the number of authors who have published in Low-Code, Modeling, and both fields. 
We also identified the most productive and influential authors within their fields and compared them to determine if both areas shared those important authors. 
An author's productivity was determined by the number of publications in the field being analyzed. 
An author's influence was determined by their h-index within the analyzed field.

Finally, to answer RQ2.5, we simply counted and listed the, previously identified, Low-Code tools and platforms explicitly related to traditional Modeling.

\subsection{Data Visualization}
The visualization of our results was performed using \textsc{Excel}, \textsc{Matplotlib}\footnote{\url{https://matplotlib.org/}}, and \textsc{Seaborn}\footnote{\url{https://seaborn.pydata.org/}}. 
To this end, we explored a variety of charts and plots, including line charts, stacked bar charts, Venn diagrams, pie charts, heatmaps, and graphs. 
Some of the visualizations were inspired by the standard ones from \textsc{Lens.org}, but we adapted them to account for our data cleaning and validation steps.

\subsection{Threats to Validity}
Our study is subjected to several threats to validity, namely: (1) internal validity, which is related to the inferences we made; and (2) external validity, which discusses the generalization of our findings.

Regarding the internal validity, Modeling venues not retrieved using our methodology, and publications related to Low-Code and Modeling paradigms that did not include the searched terms in their title, abstract, keywords, or field of study were not considered in our analysis.
Another threat relates to the curation of the collected publications.
The cleaning, validation, and categorization of Low-Code publications were performed manually, thus potentially leading to misclassified articles and metadata.
On the other hand, Modeling publications were not cleaned and validated, which may have left non-relevant publications in the datasets.
Finally, authors are identified solely by their names, so authors indexed with different names in distinct publications are treated as separate individuals in our analysis. 
Similarly, different authors indexed by the same name are considered the same person. 

As for the external validity, our results are based on data collected from \textsc{Lens.org} in October 2024, and both \textsc{GitHub} and \textsc{DBLP} in November 2024. 
Therefore, our results may not represent the entire history of publications and repositories related to Low-Code and Modeling paradigms.
Furthermore, our results should not be directly generalized to other types of publications without proper comparison and validation.

\section{Results}
\label{sec-results}

In this section, we present the results obtained during our data analysis to address the research questions identified\footnote{The data and visualizations are publicly available in \cite{tosi_2025_14808375}}.

% \subsection{Results on the Composition and Growth of the Low-Code Field}

\subsection{RQ1.1: Evolution of the Number of Low-Code Publications}
\label{sec-results-rq11}
To study the evolution of Low-Code publications, we analyzed the number of publications per year that covered this topic. 
Our analysis comprises publications from 2017, when the term ``low-code'' was first mentioned in scientific documents \cite{baldwin2017hur}, to those published until 2023 and indexed by \textsc{Lens.org} until October 2024.

\begin{figure}[htb]
    \centering
    \includegraphics[width=\linewidth]{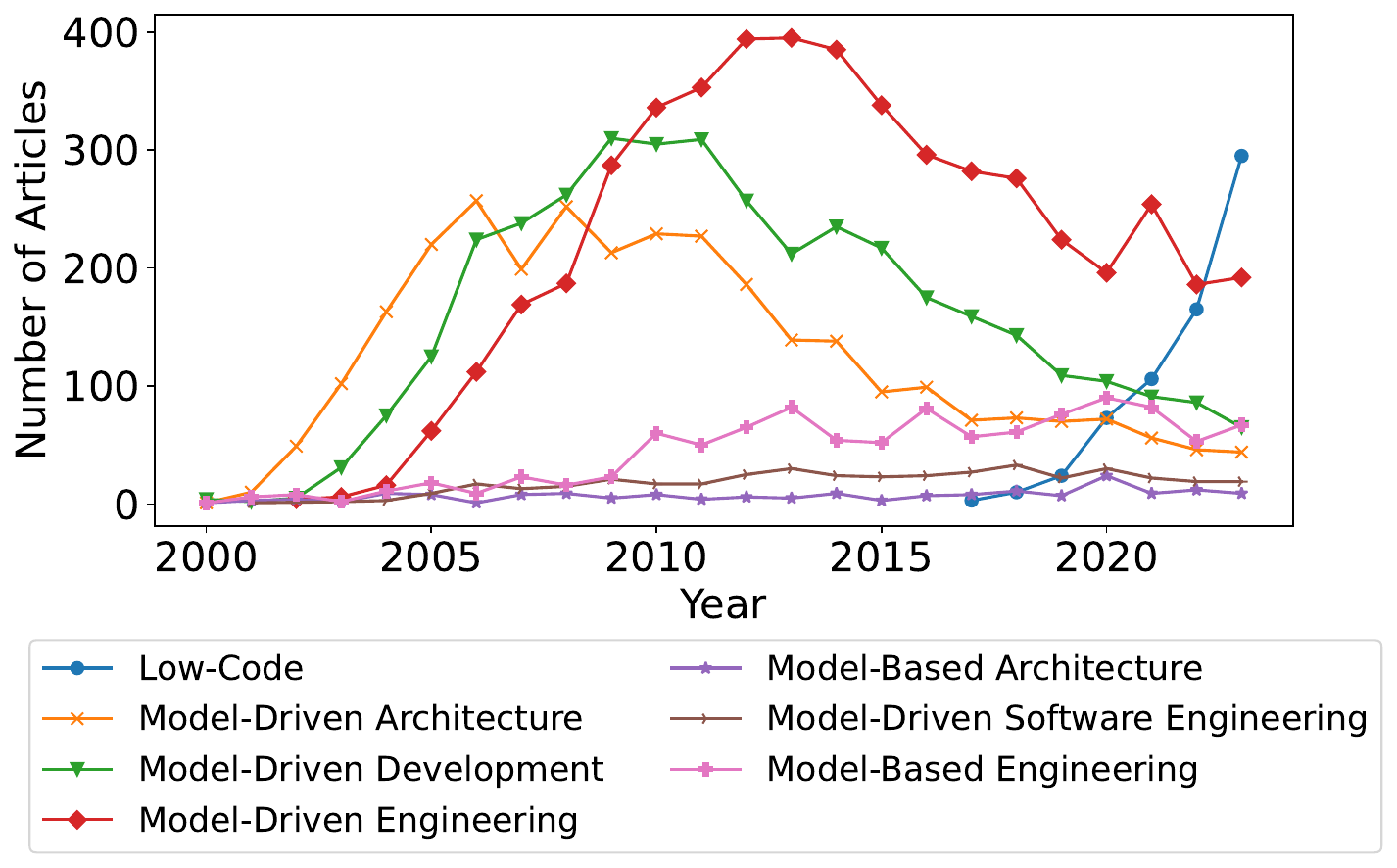}
    \caption{Publications per year involving Low-Code and traditional Modeling paradigms.}
    \label{fig:number_of_publications}
\end{figure}

Figure~\ref{fig:number_of_publications} shows the results of this analysis. 
As illustrated by the blue line, the number of Low-Code publications began to increase in 2018, with a steep rise continuing through 2023.

\begin{framed}
\noindent \textbf{Answer to RQ1.1:} 
The first Low-Code publications appeared in 2017, and their presence in the literature has been steadily increasing since then, with a marked rise in the past four years. 
\end{framed}

\subsection{RQ1.2: Low-Code Publication Profiling}
\label{sec-results-rq12}
\begin{figure}[htb]
    \centering
    \includegraphics[width=\linewidth]{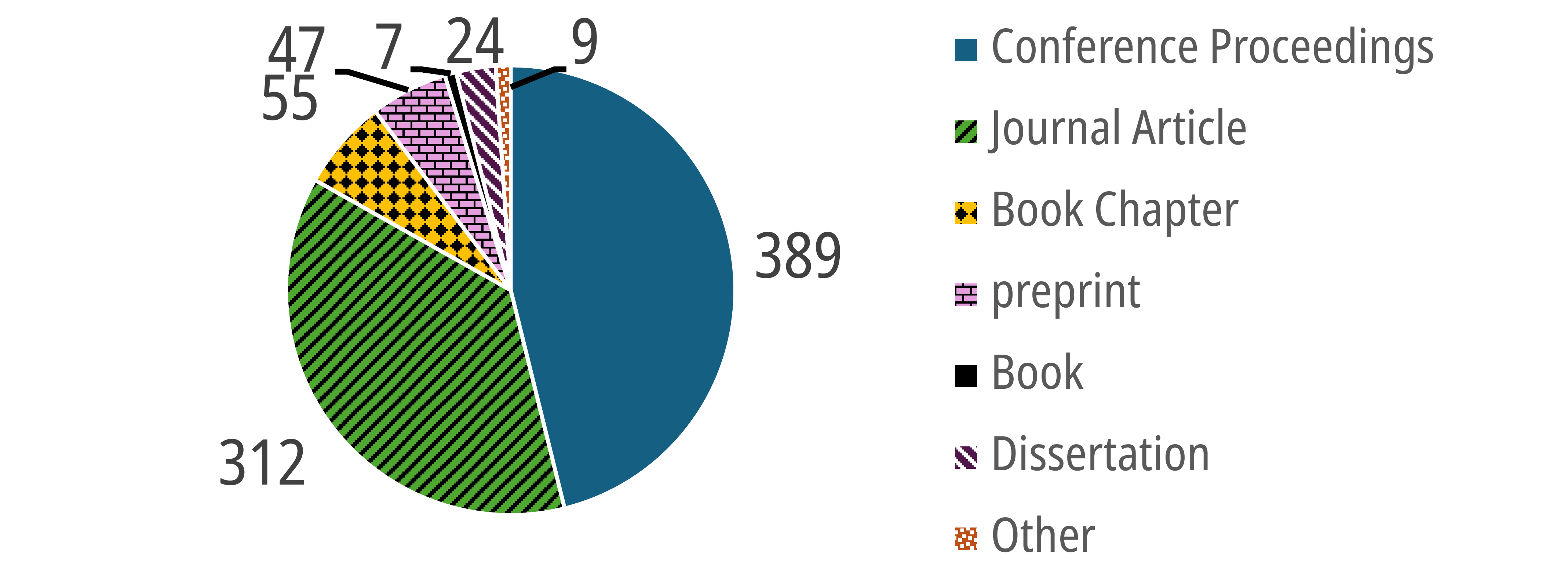}
    \caption{Types of Low-Code publications with their corresponding publication counts.}
    \label{fig:publications_type}
\end{figure}

To gain a better understanding of where Low-Code articles are being published, we present in Figure~\ref{fig:publications_type} the distribution of Low-Code publications by venue type.
Out of the 844 Low-Code publications analyzed, 46\% were published in conference proceedings (including workshops and symposiums), followed by 37\% in academic journals. The remaining 17\% were distributed across books, preprints, dissertations, and other types of publications.
Notably, only 47 Low-Code articles (less than 6\%) are preprints that have not been officially published, which are mostly stored on \textsc{arXiv} but are also available on \textsc{Zenodo} and other preprint servers. 
Of these 47 preprints, only 16 were made available before 2023.\looseness-1  

\begin{figure}[htb]
    \centering
    \includegraphics[width=\linewidth]{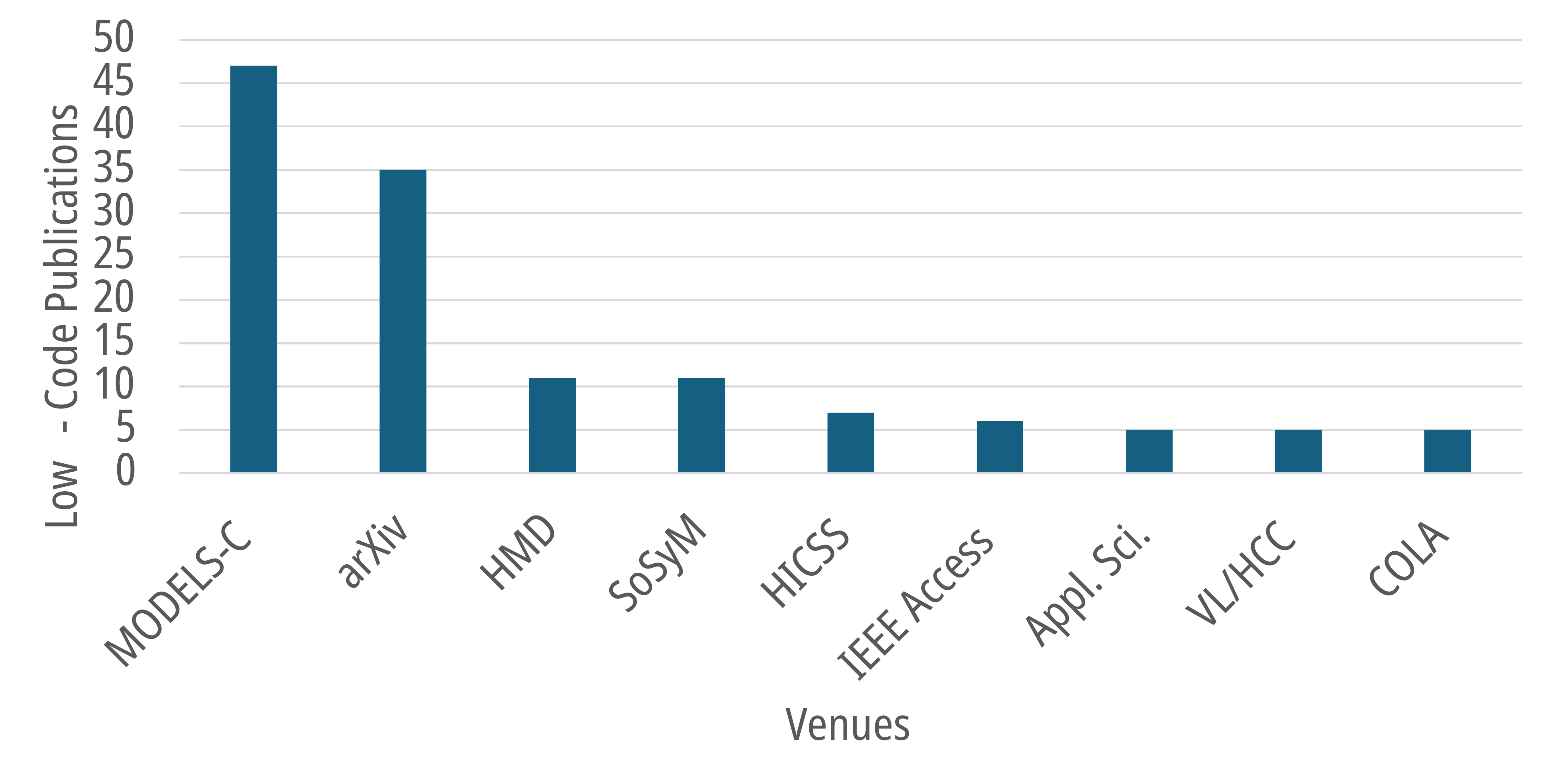}
    \caption{Top 10 venues with the highest number of Low-Code publications.}
    \label{fig:best_venues}
\end{figure}

Next, we analyze the venues that publish the most Low-Code articles. Figure~\ref{fig:best_venues} illustrates the top 10 venues based on publication counts. 
The most popular venue for Low-Code articles is the ACM/IEEE International Conference on Model-Driven Engineering Languages and Systems Companion (MODELS-C), which has almost four times as many Low-Code papers as other formal publication venues. 
MODELS-C is the proceedings dedicated to workshops and collocated tracks within the MODELS conference, and therefore, it does not represent a single venue, but the collection of satellite events associated to Models, including the International Workshop on Modeling in Low-Code Development Platforms (LowCode). 
Following MODELS-C, \textsc{arXiv} holds 35 publications, followed by the HMD Praxis der Wirtschaftsinformatik journal (HMD Practice of Business Informatics), the Software and Systems Modeling journal, the Annual Hawaii International Conference on System Sciences, and the IEEE Access journal.
Beyond the analyzed venues, Low-Code articles are dispersed across a long tail of publications, with 477 venues having published only a single Low-Code paper.

\begin{framed}
\noindent \textbf{Answer to RQ1.2:} 
Low-Code publications are mostly published in conference proceedings, accounting for 46\% of the total articles.
The ACM/IEEE International Conference on Model-Driven Engineering Languages and Systems Companion (MODELS-C) is the most popular venue for Low-Code articles, with nearly four times as many Low-Code papers as other formal publication venues. On the other hand, 56\% of Low-Code papers were published in conferences with no other Low-Code publications, suggesting that authors may lack an established venue to publish their work.
\end{framed}

\subsection{RQ1.3: Low-Code Publication Contribution Analysis}
\label{sec-results-rq13}

\begin{figure}[htb]
    \centering
    \includegraphics[width=\linewidth]{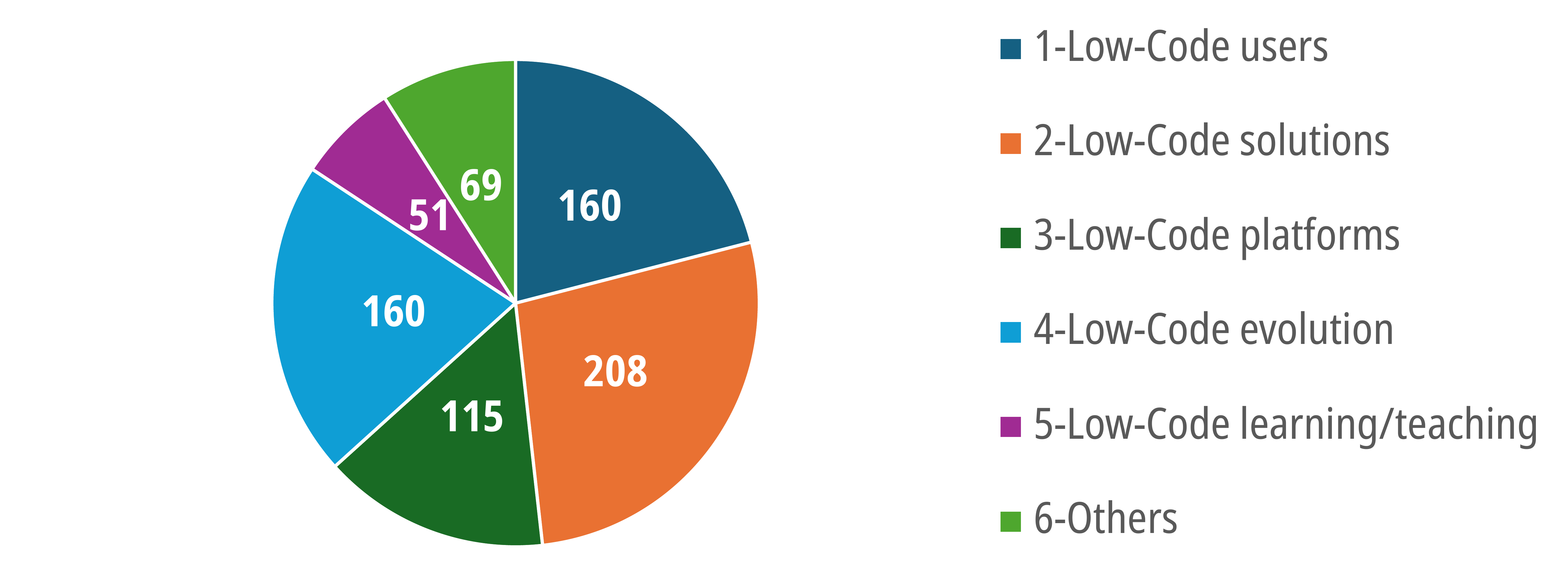}
    \caption{How publications mentioning Low-Code contribute to the research field.}
    \label{fig:contributions}
\end{figure}

To understand how Low-Code publications contribute to the research field, we manually classified the publications into six categories, as described in Section~\ref{sec-methods:rq1}.
Figure~\ref{fig:contributions} illustrates the distribution of publications across these categories.
The category with the highest number of publications is \emph{Low-Code solutions}, followed by an equal number of publications in the categories \emph{Low-Code users} and \emph{Low-Code evolution}.
These results reveal a high number of papers analyzing existing Low-Code publications (i.e., \emph{Low-Code evolution}) and expanding the use of Low-Code to solve specific problems (i.e., \emph{Low-Code solutions}).
On the other hand, we found a higher number of papers using current Low-Code tools (i.e., \emph{Low-Code users}) rather than proposing new solutions (i.e., \emph{Low-Code platforms}).
Additionally, we observed high interdisciplinarity of the Low-Code publications, which, while primarily related to Computer Science, are often applied in diverse scientific fields such as Medicine~\cite{macri2022automated}, Meteorology~\cite{xu2022design}, Business~\cite{razak2024enhancing}, among others~\cite{garas2024data, frade2022openehr}

\begin{framed}
\noindent \textbf{Answer to RQ1.3:} 
Most Low-Code publications are aimed at expanding the current state of the practice by developing Low-Code solutions for specific problems. 
Furthermore, publications presenting new Low-Code platforms are lower than those using them. Interestingly, Low-Code papers often use the paradigm as a tool in disciplines outside Computer Science, such as Medicine, Meteorology, and Business.
\end{framed}

\subsection{RQ1.4: Authorship Analysis}
\label{sec-results-rq14}

\begin{figure}[htb!]
    \centering
    \includegraphics[width=\linewidth]{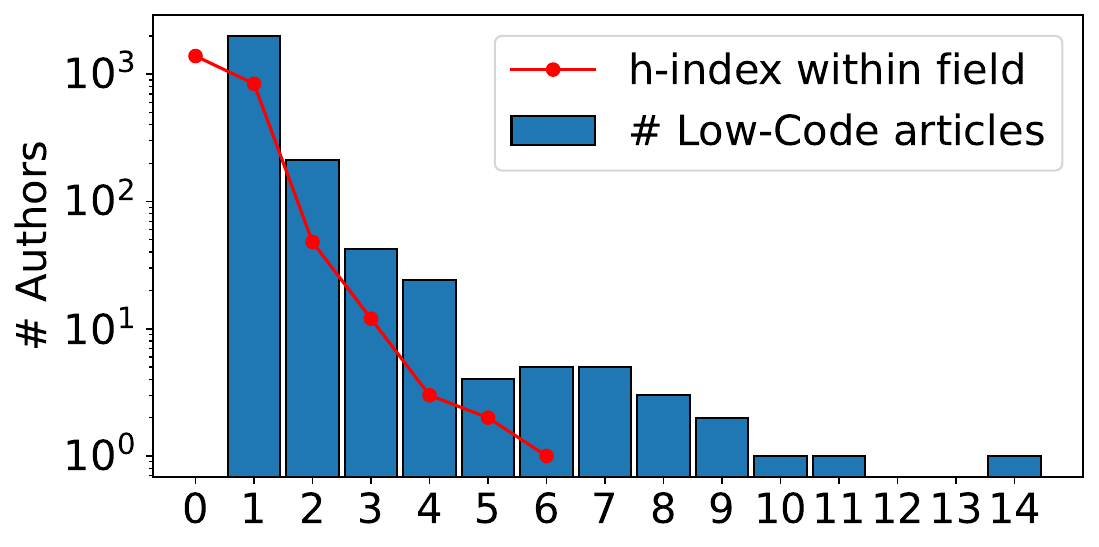}
    \caption{Number of authors with specific Low-Code publications and their h-index within the Low-Code field.}
    \label{fig:author_hindex}
\end{figure}

To investigate the characteristics of Low-Code authors, we analyzed the number of publications per author and their h-index within the Low-Code field.
As expected, the results shown in Figure \ref{fig:author_hindex} reveal a power law distribution in both the number of publications and the authors' h-index. 
With this, more than 96\% of Low-Code authors have fewer than three publications on the topic.

\begin{figure*}[htb]
    \centering
    \includegraphics[width=0.6\linewidth]{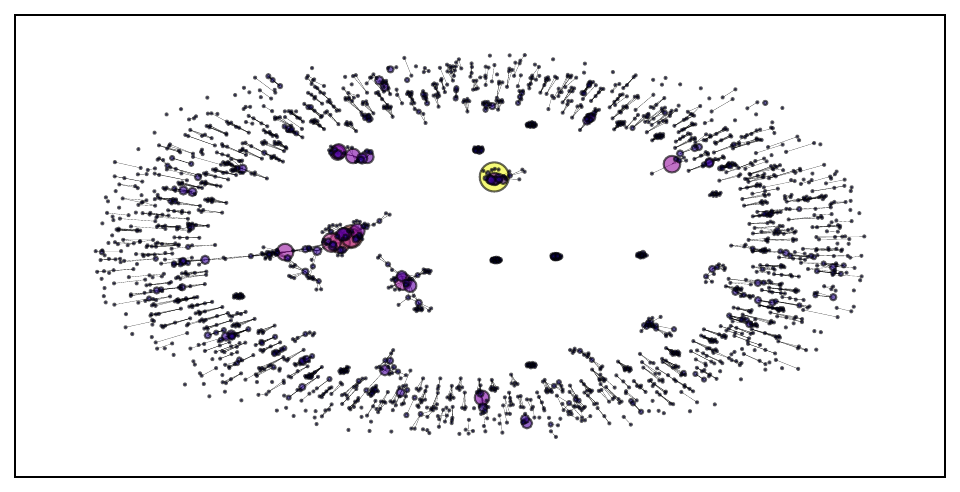}
    \caption{Co-authorship graph of Low-Code authors. Nodes and edges are positioned, colored, and sized based on the number of publications of authors and papers co-authored together respectively. Due to space limitations, closely-related nodes may overlap.}
    \label{fig:coauthors}
\end{figure*}

\begin{figure*}[htb]
    \centering
    \includegraphics[width=0.6\linewidth]{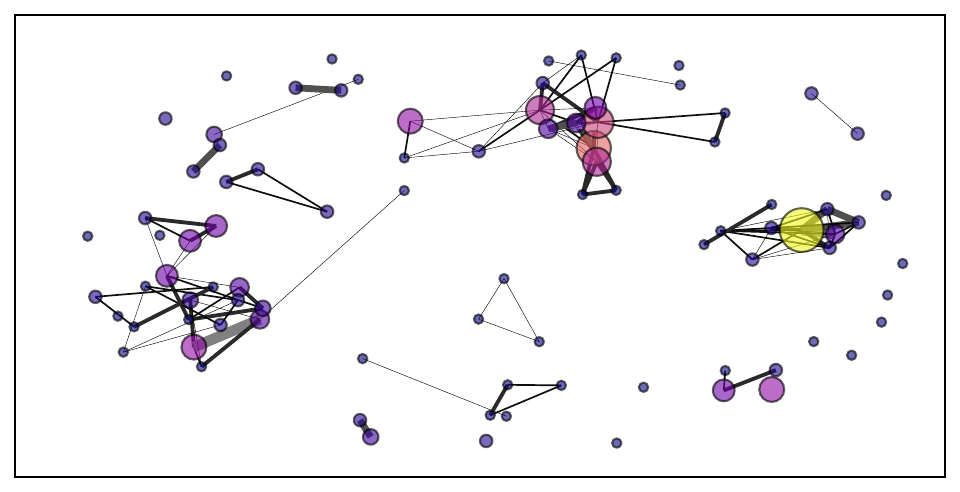}
    \caption{Co-authorship graph of Low-Code authors with at least two publications related to Low-Code. Nodes and edges are positioned, colored, and sized based on the number of publications of authors and papers co-authored together respectively. Due to space limitations, closely-related nodes may overlap.}
    \label{fig:coauthors-cleaned}
\end{figure*}

Additionally, Figure \ref{fig:coauthors} illustrates the co-authorship graph of the Low-Code authors. 
Authors with fewer publications and co-authors are shown on the outskirts of the graph, while larger co-authorship subgraphs appear in the center, primarily around researchers with more publications in the Low-Code field (represented by larger and brighter nodes). 
To better highlight the communities of active researchers, Figure~\ref{fig:coauthors-cleaned} shows the co-authorship graph focusing on authors who have published more than one paper on Low-Code (13\% of the authors). 
This visualization reveals a few larger research groups with a higher volume of Low-Code publications. Additionally, it is also possible to observe relatively low collaboration among research groups.

\begin{framed}
\noindent \textbf{Answer to RQ1.4:} 
The majority of Low-Code authors are relatively new to the field, with only 4\% of them having three or more publications related to Low-Code. 
Furthermore, authors with a larger number of publications tend to be part of larger co-authorship groups, having collaborated with a broader network of researchers. Still, collaboration among research groups is limited.
\end{framed}

\subsection{RQ1.5: Low-Code Repositories Analysis}
\label{sec-results-rq15}

To assess the popularity of Low-Code among developers, we examined open-source Low-Code tools and platforms available on \textsc{GitHub}. 
Table \ref{tab:tools} presents the 10 of 301 most-starred \textsc{GitHub} repositories that identify themselves as Low-Code. 
Notably, the table highlights the immense popularity of the topic among developers, with the most popular repository surpassing 50,000 stars. 
Additionally, we found that four of these repositories --- AppFlowy, langflow, n8n, and Flowise --- incorporate AI as part of their workflow, indicating a growing trend of integrating artificial intelligence (AI) in Low-Code solutions.

\begin{table*}[htb]
    \centering
    % \fontsize{6pt}{6pt}\selectfont
    \footnotesize
    \begin{tabularx}{\textwidth}{lcX}
        \multicolumn{1}{c}{\textsc{Name}} & \multicolumn{1}{c}{\textsc{Stars}} & \multicolumn{1}{c}{\textsc{Description}} \\
        \toprule
        AppFlowy & 58,145 & Bring projects, wikis, and teams together with AI. AppFlowy is an AI collaborative workspace where you achieve more without losing control of your data. The best open source alternative to Notion. \\
        nocodb   & 49,791  & Open Source Airtable Alternative \\
        n8n      & 49,107 &  Free and source-available fair-code licensed workflow automation tool. Easily automate tasks across different services. \\
        langflow & 34,668 & Langflow is a low-code app builder for RAG and multi-agent AI applications. It's Python-based and agnostic to any model, API, or database. \\
        appsmith & 34,653 & Platform to build admin panels, internal tools, and dashboards. Integrates with 25+ databases and any API.  \\
        ToolJet  & 33,002 & Low-code platform for building business applications. Connect to databases, cloud storages, GraphQL, API endpoints, Airtable, Google sheets, OpenAI, etc and build apps using drag and drop application  \\
        Flowise  & 31,653 & Drag \& drop UI to build your customized LLM flow  \\
        refine   & 28,564 & A React Framework for building  internal tools, admin panels, dashboards \& B2B apps with unmatched flexibility.  \\
        budibase & 22,749 & Low code platform for building business apps and workflows in minutes. Supports PostgreSQL, MySQL, MariaDB, MSSQL, MongoDB, Rest API, Docker, K8s, and more  \\
        node-red & 19,915 & Low-code programming for event-driven applications  \\
        \bottomrule

    \end{tabularx}
    \caption{Most popular Low-Code tools and platforms on \textsc{GitHub}.}
    \label{tab:tools}
\end{table*}

\begin{framed}
\noindent \textbf{Answer to RQ1.5:} 
There is a large number of Low-Code repositories on \textsc{GitHub}, many of which have garnered significant attention from the community, achieving tens of thousands of stars. 
This demonstrates that the Low-Code topic is being actively explored by developers. 
\end{framed}

% \subsection{Results on the Impact of Low-Code in the Modeling community}

\subsection{RQ2.1: Impact of Low-Code in Modeling publications}
\label{sec-results-rq21}

To examine the impact of Low-Code on traditional Modeling publications, we analyzed the number of publications per year that involved one (or the two) paradigms.
In addition to Low-Code, we focused on the following traditional Modeling paradigms: Model-Driven Architecture (MDA), Model-Driven Development (MDD), Model-Driven Engineering (MDE), Model-Based Architecture (MBA), Model-Driven Software Engineering (MDSE), and Model-Based Engineering (MBE).
The analysis covered publications from 2000 to 2023.

As shown in Figure \ref{fig:number_of_publications}, the number of papers on traditional Modeling paradigms grew steadily from 2000 to around 2012 but began to decline thereafter. 
MDA papers, followed by MDD and MDE, with MDE showing the longest presence in the literature.
Note that the number of papers focusing on MBA, MDSE, and MBE has always been relatively low for the period analyzed.
During the decrease in traditional Modeling papers, the number of Low-Code papers started to increase in 2018, with a steep rise until 2023, the last year of our analysis.
In 2023, the presence of Low-Code papers has been higher than the presence of any other individual modeling-related term.

\begin{figure}[htb]
    \centering
    \includegraphics[width=0.92\linewidth]{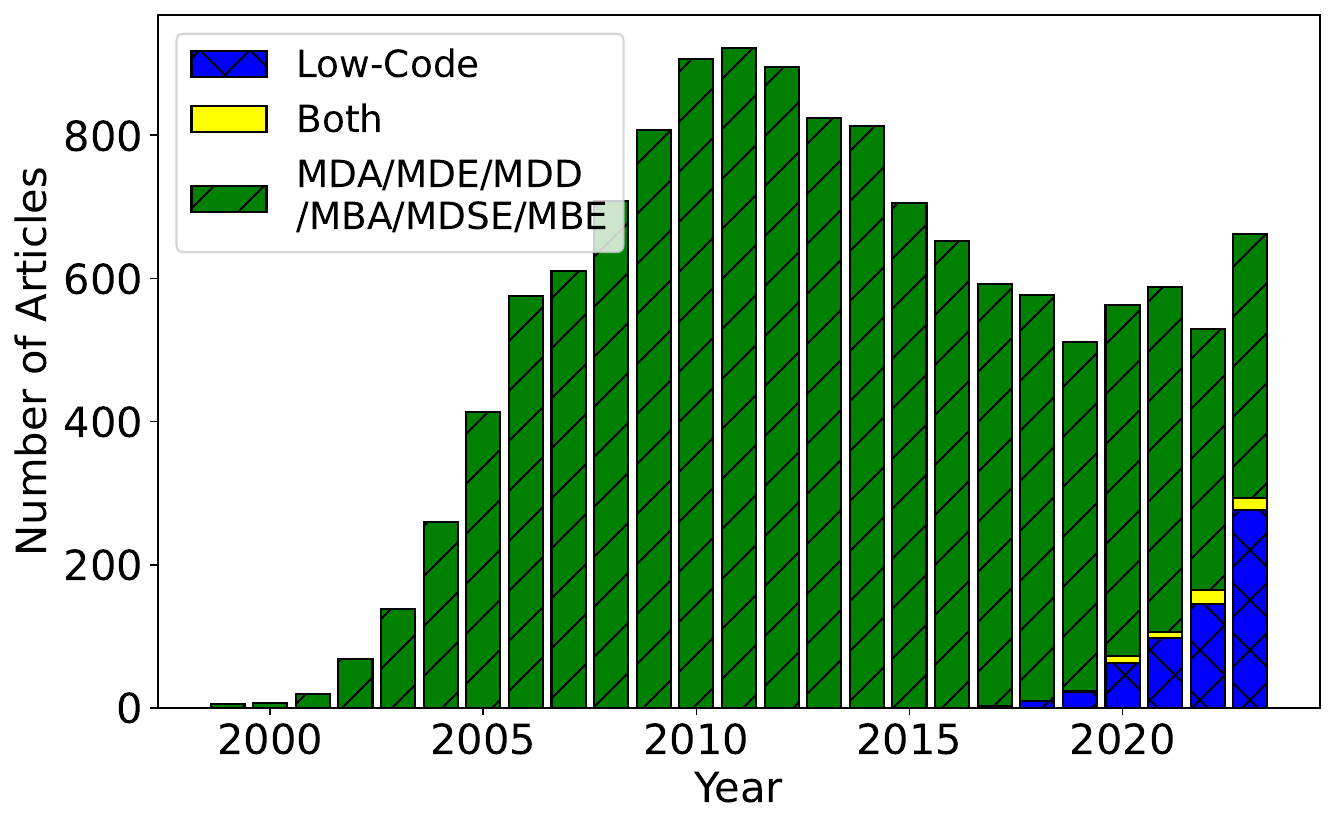}
    \caption{Combined publications per year involving Low-Code and traditional Modeling paradigms.}
    \label{fig:lowcode_vs_modeling}
\end{figure}

To better compare the evolution of Low-Code and traditional Modeling papers, Figure~\ref{fig:lowcode_vs_modeling} presents the number of publications per year for three main groups: (1) Low-Code papers, (2) traditional Modeling papers, and (3) papers involving both paradigms.
As observed, the number of Low-Code papers in 2023 (294 papers) is already close to the number of traditional Modeling papers (385 papers). 
On the other hand, the number of papers involving both paradigms is still negligible. 

\begin{framed}
\noindent \textbf{Answer to RQ2.1:} 
Low-Code publications began to increase about 6 years after traditional Modeling publications started to decline. Today, Low-Code publications receive comparable attention from researchers as to that of traditional Modeling paradigms. 
\end{framed}

\subsection{RQ2.2: Modeling conferences Targeting Low-Code}
\label{sec-results-rq22}

To explore the extent to which Modeling venues are targeting Low-Code, we compiled a list of 64 Modeling conferences and 84 workshops. 
In Table \ref{tab:venues-target}, we present the venues that mention ``Low-Code'' in their website to actively attract researchers and enthusiasts to the topic. 
From the table, we see that 7 conferences and 7 workshops explicitly target Low-Code. Three of each focus on Low-Code as a specific topic of interest, while the remaining mention Low-Code alongside other keywords without giving it much emphasis. 
Notably, the MODELS conference stands out for giving significant focus to Low-Code. 
Although the conference itself does not prioritize Low-Code as a core topic, it hosts three workshops that do, including the International Workshop on Modeling in Low-Code Development Platforms (LowCode), the only venue created specifically to address Low-Code research.

\begin{table*}[htb]
    \renewcommand{\arraystretch}{0.6}
    \centering
    \footnotesize
    \begin{tabularx}{\textwidth}{ccccX}
        \textsc{Venue}& \multirow{2}{*}{\textsc{Acronym}} & \textsc{Low-Code} & \textsc{Type of} & \multirow{2}{*}{\textsc{Excerpt from the venue website}} \\
        \textsc{Type} &  & \textsc{Focus} & \textsc{Focus} &  \\
        \toprule
        \multirow{8}{*}{\rotatebox[origin=c]{90}{Conference\hspace{6.5em}}}
         & MODELS & No & Topics of interest & New paradigms, formalisms, applications, approaches, frameworks, or processes for model-based engineering such as \textbf{low-code}/no-code development, digital twins, etc. \\
        \cmidrule{2-5}
         %& MODELSWARD & No & Keynote & The Secret Recipe to the Perfect \textbf{Low-Code} Platform. \\
         & ECMFA & No & Topics of interest & New paradigms, formalisms, applications, approaches, frameworks, or processes for model-based engineering such as \textbf{low-code}/no-code development, digital twins, etc. \\
        \cmidrule{2-5}
         & PoEM & Yes & Topics of interest & \textbf{LowCode} NoCode tools. \\
        \cmidrule{2-5}
         & EMSOFT & No & Tutorial & \textbf{Low Code}, High Performance Embedded AI with MATLAB \& Arm IP Explorer.\\
         \cmidrule{2-5}
         & SAM & No & Call for Papers & The scope included advancements/usage of languages/methods standardized by the ITU-T as well as domain-specific languages for state-of-practice domains like artificial intelligence, digital twins, no code, \textbf{low code}, DevOps, and metaverse. \\
         \cmidrule{2-5}
         & MDASD & Yes & Topics of interest & \textbf{Low-Code} and No-Code software development – research, experiences and challenges.\\
         \cmidrule{2-5}
         & MDEML & Yes & Motivation & Modeling languages and MDE are currently leveraged in most applicative domains, from banking to digital printing, and drive the success story of \textbf{low-code}, no-code and other development paradigms.\\
         \cmidrule{4-5}
         &  &  & Topics of interest & \textbf{Low-code} and/or no-code applications. \\
         %& STAF & No & Keynote &  \textbf{Low-Code} and Low-Modeling Strategies for Agile MDE Processes. \\
         \midrule
         \multirow{8}{*}{\rotatebox[origin=c]{90}{Workshop\hspace{5em}}}
         & VisMod@MODELS & Yes & Topics of interest & \textbf{LowCode}/NoCode techniques and methodologies. \\
         \cmidrule{2-5}
         & MoDRE@RE & Yes & Topics of interest & Requirements engineering approaches for \textbf{low-code}/no-code software development. \\
         \cmidrule{2-5}
         & LowCode@MODELS & Yes & Objectives & Bring together developers and users of \textbf{low-code} platforms with model-driven engineering researchers and practitioners; ...\\
         \cmidrule{4-5}
         & & & Topics of interest & Technologies underpinning \textbf{low-code} platforms; ...\\
         \cmidrule{2-5}
         & FPVM@MODELS & Yes & Topics of Interest & \textbf{LowCode}/NoCode techniques and methodologies. \\
         \cmidrule{2-5}
         & HybridAIMS@CAiSE & No & Topics of Interest & \textbf{Low code} approaches for, e.g., Knowledge Graphs, Machine Learning, knowledge engineering, Hybrid AI engineering. \\
         \cmidrule{2-5}
         & KG4SDSE@CAiSE & No & Description & We also aim to investigate (...) how Knowledge Graphs enable new flavors of model-driven engineering or \textbf{low-code} engineering... \\
         \cmidrule{2-5}
         & LLM4MDE@STAF & No & Topics of Interest & LLMs for supporting \textbf{low-code} development\\
         \bottomrule

    \end{tabularx}
    \caption{Modeling venues targeting Low-Code.}
    \label{tab:venues-target}
\end{table*}

\begin{framed}
\noindent \textbf{Answer to RQ2.2:} 
Few Modeling venues are specifically targeting Low-Code. 
Despite the increasing attention Low-Code is receiving from researchers and developers, the number of Modeling events encouraging submissions on this topic remains limited.
\end{framed}

\subsection{RQ2.3: Presence of Low-Code in Modeling Conferences}
\label{sec-results-rq23}

Given the growing attention Low-Code has been receiving from researchers, as shown in Section~\ref{sec-results-rq11}, we investigate the presence of the topic in the papers published in three Modeling venues: ECMFA, MODELS, and MODELS-C --- each of which specifically targets Low-Code as part of their topics of interest.
The results of this investigation are summarized in Table~\ref{tab:keywords}. 
As expected, the upper part of the table highlights the most prominent concepts related to traditional Modeling, such as ``model'', ``DSL'', and ``language''. 
On the other hand, on the bottom part of the table, we observe that Low-Code related keyphrases are not considered highly relevant, indicating lower presence in the analyzed venues. 
Notably, the term ``low-code'' only appeared as a separate keyphrase in the MODELS-C workshops, and even there, it ranked 51st.

\begin{table*}[htb]
    \centering
    \footnotesize
    \begin{tabularx}{0.7\textwidth}{ccccccc}
        \multirow{2}{*}{\textsc{Rank}} & \multicolumn{2}{c}{\textsc{MODELS}} & \multicolumn{2}{c}{\textsc{MODELS-C}} & \multicolumn{2}{c}{\textsc{ECMFA}} \\ 
        & \textsc{Concept} & \textsc{Weight} & \textsc{Concept} & \textsc{Weight} & \textsc{Concept} & \textsc{Weight} \\
        \toprule
        1 & model& 1.0 &model& 1.0 &model& 1.0 \\
        2 & approach& 0.280 &system& 0.546 &system& 0.419 \\ 
        3 & software& 0.271 &tool& 0.453&approach& 0.322\\
        4 & system& 0.271 &modeling& 0.379&engineering& 0.225\\
        5 & based& 0.233 &approach& 0.364&information& 0.209\\
        6 & data& 0.205 &language& 0.250&development& 0.177\\
        7 & development& 0.196 &challenge& 0.226&paper& 0.177\\
        8 & tool& 0.186 &domain& 0.223&support& 0.177\\
        9 & design& 0.186 &development& 0.217&based& 0.177\\
        10 & paper& 0.186 &data& 0.214&view& 0.177\\
        11 & domain& 0.186 &different& 0.208&model driven& 0.161\\
        12 & using& 0.177 &support& 0.208&domain& 0.145\\
        13 & modeling& 0.158 &based& 0.2&DSL& 0.145\\
        14 & Alloy& 0.158  &paper& 0.2&infrastructure& 0.145\\
        15 & prompt& 0.140 &software& 0.2&CD& 0.145\\
        \bottomrule
        \textsc{Concept} & \multicolumn{2}{c}{\textsc{MODELS}} & \multicolumn{2}{c}{\textsc{MODELS-C}} & \multicolumn{2}{c}{\textsc{ECMFA}} \\
        & \textsc{Rank} & \textsc{Weight} & \textsc{Rank} & \textsc{Weight} & \textsc{Rank} & \textsc{Weight} \\
        \hline
         \textbf{low} & 134 & 0.046 & - & - & - & - \\
         \textbf{code} & 17 & 0.130 & 68 & 0.077 & 74 & 0.0645 \\
         \textbf{low code}& - & - & 51 & 0.095 & - & - \\

    \end{tabularx}
    \caption{Rank of most relevant keyphrases in MODELS, MODELS-C and ECMFA in 2023.}
    \label{tab:keywords}
\end{table*}

\begin{framed}
\noindent \textbf{Answer to RQ2.3:} 
Low-Code does not appear as one of the most prominent topics in any of the proceedings of the Modeling venues we analyzed. 
This suggests that the Low-Code topic has not yet fully integrated into the traditional Modeling community despite the openness of such conferences. 
\end{framed}

\subsection{RQ2.4: Authorship Distribution between Low-Code and Modeling}
\label{sec-results-rq24}

\begin{figure*}[htb]
\centering
\begin{subfigure}{.45\linewidth}
  \centering
  \includegraphics[width=0.6\columnwidth]{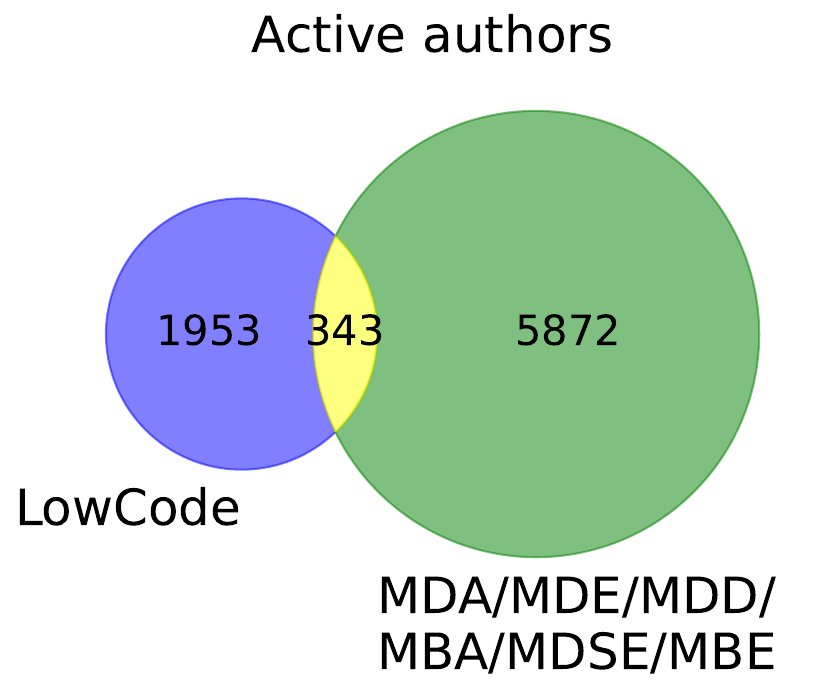}
    \caption{}
    \label{fig:active_authors}
\end{subfigure}%
\begin{subfigure}{.45\linewidth}
  \centering
  \includegraphics[width=0.6\columnwidth]{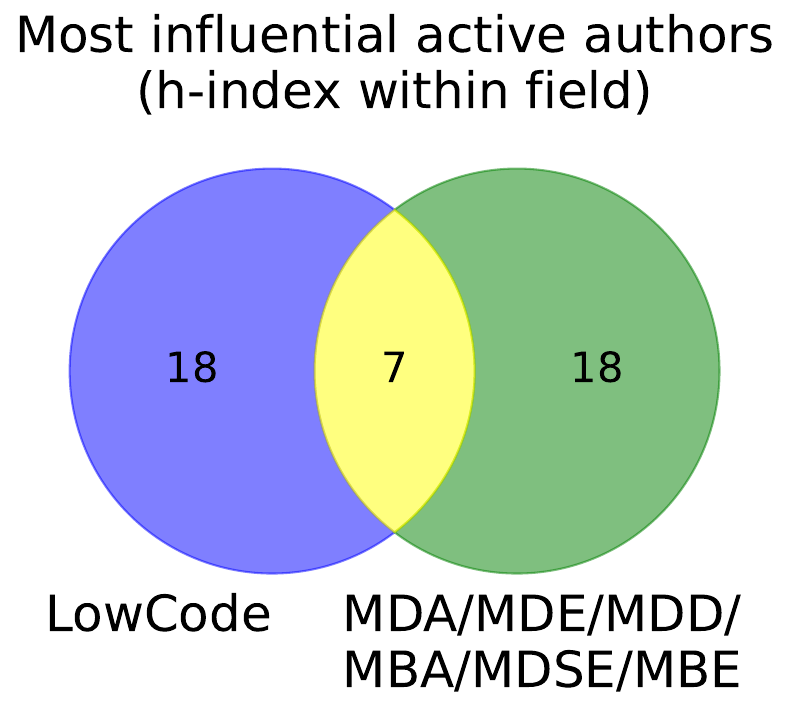}
    \caption{}
    \label{fig:most_influential_authors}
\end{subfigure}
\caption{Authors with publications involving Low-Code and traditional Modeling from 2018 onward.}
\label{fig:authors}
\end{figure*}

To study the distribution of authors between Low-Code and traditional Modeling, we first identified all authors who have been active in these areas. 
We define an author as active if they have at least one publication in either Low-Code or Modeling since 2018.
Figure~\ref{fig:active_authors} illustrates the distribution of active authors in the studied areas. 
As shown, the overlap between authors who have published in both traditional Modeling and Low-Code is relatively small. 
Specifically, only around 15\% of Low-Code authors have also authored a publication related to traditional Modeling.  
This may be due to the interdisciplinarity of Low-Code papers, observed in Section \ref{sec-results-rq13}. 

\begin{figure}[htb]
    \centering
    \includegraphics[width=0.73\linewidth]{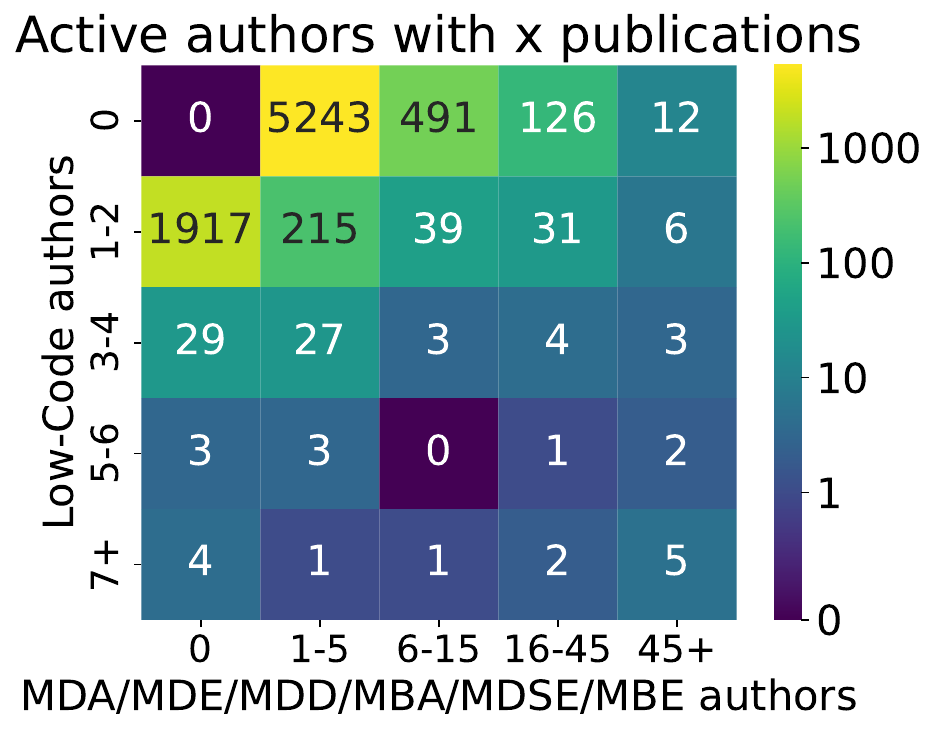}
    \caption{Heatmap of active authors by number of publications in Low-Code and traditional Modeling paradigms.}
    \label{fig:heatmap}
\end{figure}

After analyzing all active authors, we focused on comparing only the most productive and influential authors. 
Figure~\ref{fig:most_influential_authors} shows the overlap of the 18 most influential authors in Low-Code and traditional Modeling. 
Higher than in the previous analysis, we can observe 28\% of overlap between the most influential authors in Low-Code and Modeling. 
Additionally, Figure~\ref{fig:heatmap} illustrates the distribution of authors based on their productivity in Low-Code and traditional Modeling.  

From Figure \ref{fig:heatmap}, we observe that 45\% of the 22 most productive Low-Code authors (those with at least 5 publications, as shown in the last two rows of the distribution matrix) are also highly productive in the traditional Modeling community (with at least 16 publications, as indicated in the last two columns of the last two rows of the distribution matrix).
On the other hand, we also see that 7 of these 22 authors have no publications in traditional Modeling (indicated in the first column of the last two row).

\begin{framed}
\noindent \textbf{Answer to RQ2.4:} 
The overlap between authors who have published in both traditional Modeling and Low-Code is small, around 15\% of the analyzed authors.
Furthermore, 10 of the most influential and productive authors in Low-Code also have a significant presence in the traditional Modeling community.
On the other hand, some authors are highly productive in Low-Code but have no publications in traditional Modeling.
\end{framed}

\subsection{RQ2.5: Low-Code tools explicitly related to Modeling}
\label{sec-results-rq25}

Table \ref{tab:model-tools} lists open-source Low-Code tools that are explicitly model-based. These tools constitute only 6\% of the 151 Low-Code tools available on GitHub according to our criteria.
From this table, we identify two primary categories of Low-Code platforms and tools explicitly related to traditional Modeling. 
The first category covers AI-related Low-Code tools --- specifically \textit{langflow}, \textit{ludwig}, and \textit{BESSER}~\cite{alfonso2024building}) --- which aim to reduce the amount of code required to develop AI-based softwares. 
The second category covers CRUD-related Low-Code tools, which focus on automating operations for creating, reading, updating, and deleting information --- specifically \textit{evolutility-ui-jquery}, \textit{evolutility-ui-react}, \textit{evolutility-server-node}, and \textit{crud}.
An additional observation is that \textit{BESSER} --- project in which two of the authors are involved --- is the only one among those platforms that provides citation instructions. 
This indicates it is the only research-oriented platform, focusing not only on increasing its adoption by developers but also contributing to Low-Code research.  

\begin{table*}[htb]
    \renewcommand{\arraystretch}{1}
    \footnotesize
    \centering
    \begin{tabularx}{\linewidth}{ccX}
        \multicolumn{1}{c}{\textsc{Name}} & \multicolumn{1}{c}{\textsc{Stars}} & \multicolumn{1}{c}{\textsc{Description}} \\
        \toprule
         langflow & 34311 & Langflow is a low-code app builder for RAG and multi-agent AI applications. It’s Python-based and agnostic to any model, API, or database. \\
         \midrule
         ludwig & 11187 & Low-code framework for building custom LLMs, neural networks, and other AI models  \\
         \midrule
         dgiot & 6502 &  Open source platform for iot , 6 min Quick Deployment,10M devices connection,Carrier level Stability Low code for Object model-Rule Engine-Data Channel-Configuration Page; Fully open source, Multi industrial protocols are compatible.\\
         \midrule
         system-designer & 936 &  A low-code development platform for creating systems.\\
         \midrule
         evolutility-ui-jquery & 183 &  Model-driven Web UI for CRUD using REST or localStorage. \\
         \midrule
         evolutility-ui-react & 113 &  Framework for building CRUD UIs for Hasura GraphQL with models rather than code. \\
         \midrule
         evolutility-server-node & 112  &  Model-driven REST APIs for CRUD and more, written in Javascript, using Node.js, Express, and PostgreSQL. \\
         \midrule
         crud & 66 &  A package helps writing CRUD servers. All you need is this package and models. \\
         \midrule
         BESSER & 56 &  A Python-based low-modeling low-code platform for smart software. \\
         \bottomrule 
    \end{tabularx}
    \caption{Low-Code tools on \textsc{GitHub} explicitly related to traditional modeling.}
    \label{tab:model-tools}
\end{table*}

\begin{framed}
\noindent \textbf{Answer to RQ2.5:} 
Only nine Low-Code platforms and tools available on GitHub are explicitly related to traditional Modeling. These tools primarily focus on accelerating the development of AI-based software or automating CRUD-related operations. Among them, \textit{BESSER} is the only research-oriented platform identified.
\end{framed}
\section{Discussion}
\label{sec-discussion}
Beyond the main results reported so far, we highlight below additional insights derived from our findings. 
These insights focus on existing and potential challenges within the Modeling and Low-Code communities. 
We also propose recommendations to address these challenges while improving the synergy between the two communities.

\begin{itemize}
    \item \textbf{Decline in the modeling community}. 
Over the past five years, the annual number of Low-Code papers has reached a value higher than the yearly number of any individual Modeling paradigm and is approaching the total number of all Modeling-related papers combined.
This phenomenon is driven not only by the high interest in Low-Code in recent years but also by a decreasing interest in traditional Modeling, especially outside the core modeling conferences.
Considering that scientific research is known for having bubble trends, Low-Code interest may have negatively impacted the number of traditional Modeling publications but we cannot guarantee such behavior is causal or not.
Thus, these shifts in publication behavior may reveal a migration pattern that requires further investigation, particularly from authors who have transitioned from MDE, the most popular traditional modeling paradigm, to Low-Code. This migration plays a key role in transferring valuable expertise from traditional modeling to the Low-Code field but also raises concerns about the potential ``reinvention of the wheel'', a topic we address in our next insight.
Furthermore, despite its distinct characteristics, Low-Code remains a modeling approach. When analyzing the overall number of modeling publications, the rise of Low-Code papers has softened the decline in total output.
But despite this significant high interest in Low-Code, the combined number of papers on Low-Code and traditional modeling still does not reach the all-time high for the Modeling community observed around 2011. 

It is unclear whether interest in low-code will continue to grow and allow the modeling community to reverse the trend by bringing more people in the community or if we will instead see a kind of shift in the topics of the papers without consolidating into the modeling community the first-time authors of low-code papers. 
Given that one factor in the decline may partially be caused by the fragmentation of the Modeling field into distinct topics (e.g. MDA, MDE, MDD, etc.), which can have harmed its adoption as the diversity of terms hinders a broader recognition of the field, centering many new papers around the low-code terminology could also increase the recognition and focus the attention of outsiders that may consider entering the community and right now are confused by the terminology.
\\[0.5\baselineskip]
\textbf{Recommendation.} Authors from both the Modeling and the Low-Code communities should recognize the complementary/overlapping nature of their fields.
Some Modeling authors may be unfamiliar with Low-Code and, consequently, treat it as it if was an unrelated field of study missing the opportunity to communicate that their modeling findings could, often, be also useful for low-code practitioners, potentially decreasing their influence and overall impact of their contributions. 
%Therefore, we believe that traditional Modeling authors should be aware that Modeling contributions can be directly -- or with minimum adaptations -- applied to Low-Code solutions. 
Indeed, presenting their solutions from a Low-Code perspective can improve adoption rates and the size of its target audience. %, ultimately increasing the influence and impact of their contributions. 
On the opposite side, the Low-Code community can benefit from decades of foundation research and experience from the traditional Modeling community.
This expertise can help the Low-Code field to overcome challenges already addressed or mitigated by the Modeling community. By becoming part of this larger Modeling community they could benefit from this preexisting know-how and accelerate the scientific progress of the Low-code field. 
\\
    \item \textbf{Do not reinvent the wheel}. 
The fact that there is a 28\% of overlap between the most influential authors in Low-Code and Modeling may reveal that senior authors from the Modeling community are adopting the Low-Code terminology and techniques, and, consequently, influencing younger co-authors. 
Still, approximately 85\% of Low-Code authors lack prior experience in traditional modeling research (cf. Figure~\ref{fig:heatmap}). 
\\[0.5\baselineskip]
\textbf{Recommendation.} Senior researchers should guide the younger members of the community about the usefulness of knowing existent modeling publications and their terminologies to avoid reinventing the wheel.
Moreover, we emphasize that a low-code paper that just reinvents an algorithm/technique already available in the modeling community and just uses a low-code terminology as key difference is not acceptable. 
\\
    \item \textbf{Workshops and \textsc{arXiv} as entry points for Low-Code}. 
Our results indicate that Low-Code is entering the Modeling community primarily through workshops, particularly those associated with the MODELS conference (cf. Section \ref{fig:best_venues}). 
We believe that this behavior aligns with typical scientific publication practices, where novel applications, usually from other domains, approach new disciplines, but it requires further validation. 
Additionally, the significant presence of publications on \textsc{arXiv} suggests new trends in research dissemination. 
Consequently, as the topic continues to attract attention from researchers both within and beyond the Modeling community, we anticipate an increase in Low-Code publications being presented at major conferences and journals.
\\[0.5\baselineskip]
\textbf{Recommendation.} The modeling community should keep an eye on the situation and make sure there are no biases against low-code papers that could prevent them from entering the core research tracks in the conferences instead of just being relegated to satellite events. 
\\
    \item \textbf{Low-Code papers outside of Modeling venues}. 
Currently, the majority of Low-Code papers are published in venues unrelated to traditional Modeling, despite the significant reuse of Modeling concepts within Low-Code (Section~\ref{sec-results-rq12}). 
Furthermore, 56\% of the Low-Code papers were published on venues that have not published any other Low-Code publication, which may indicate Low-Code papers may lack an established venue in the area. 
These trends may come from the limited emphasis placed by most modeling venues on encouraging Low-Code-related tracks and submissions (Section~\ref{sec-results-rq22}).
We think modeling conferences should put some effort in inviting and bringing into the modeling community these authors. 
\\[0.5\baselineskip]
\textbf{Recommendation.} Traditional Modeling venues should further encourage the submission of Low-Code-related research to enhance the synergy between Low-Code and Modeling. 
Maybe as part of application tracks, tool demos or even special workshops trying to facilitate the cross-fertilization between modeling experts and domain experts. 
This could initially focus on topics that bridge the two fields, such as applying traditional modeling techniques to address Low-Code challenges or presenting novel Low-Code solutions inspired by modeling principles.
Such measures would also help creating a relationship between the core modeling conferences (where new techniques should be presented) and conferences in other domains (where modeling is a means to an end).
These initiatives could improve the visibility of traditional modeling venues among Low-Code authors, thereby narrowing the gap between the two communities. 
\\
    \item \textbf{Low industry association between Low-Code and Modeling}. 
Currently, only 6\% of Low-Code platforms and tools on GitHub are explicitly model-related (see Section~\ref{sec-results-rq25}).
This suggests that tool developers --- and by extension, the industry --- do not directly associate Low-Code with Modeling.  
\\[0.5\baselineskip]
\textbf{Recommendation.} Explain to industry that Low-Code is a subset of modeling by creating more opportunities for showcasing Low-Code solutions in traditional Modeling venues on industry tracks, demonstration sessions, and related forums. 
Such measure could encourage the exchange between both communities and additionally highlight the importance of traditional Modeling solutions to Low-Code developers. 
This larger visibility could drive broader adoption of Modeling methodologies, thus increasing their impact.
\\
    \item \textbf{Research on Low-Code Improvements}. 
During the analysis of Low-Code publications (Section~\ref{sec-results-rq13}), we observed a substantial number of \textit{Low-Code solutions}. 
However, most of these contributions are domain-specific and do not aim to improve the broader Low-Code domain or address general challenges in the field. 
Instead, they often focus on developing frameworks for specific tools or applications. 
\\[0.5\baselineskip]
\textbf{Recommendation.}
We see a clear need for research that advances the Low-Code paradigm itself.
We believe that it would be important to address topics such as the usability of Low-Code tools, best practices for developing Low-Code platforms, how to test Low-Code software properly, and the integration of existing Modeling solutions into Low-Code pipelines.
Such measure would foster the growth and evolution of the field.

\end{itemize}

\section{Conclusion}
\label{sec-conclusion}

In this paper, we have studied how Low-Code has increased in popularity over the last decade and what this means for the research community being developed over it. 
In particular, in our results, we have identified the composition of the Low-Code community, discussed important trends, and highlighted its similarities and differences compared to the traditional Modeling community. 
As a result, we observed that despite prominent researchers from the traditional Modeling community fostering Low-Code, most of Low-Code researchers and developers are not active members of the Modeling community and/or do not explicitly integrate Modeling into their solutions yet. So there is margin for improvement and cross-fertilization between the two communities. 

Ultimately, based on those observations, we propose a set of recommendations to the traditional Modeling community on how to enhance the synergy between the Low-Code and the Modeling communities. 
We expect with such results and recommendations to trigger discussions about the future of Low-Code research and help the community to identify what it wants to become as it consolidates itself. 
We plan to continue monitoring the metrics and results of our study to observe whether the trends we have identified continue to hold true and/or whether both communities are able to complement better together to maximize the size, and more importantly, the impact of the overall community. 
Given the limited impact of modeling (including low-code) in software engineering research\footnote{An anecdotal evidence could be to just count the close-to-none number of papers on core modeling topics in major software engineering conferences such as ICSE}, we want to make sure all people somehow working on modeling-related topics (in the broadest sense) contribute to push the importance of these topics.

\begin{acks}
This project is supported by the Luxembourg National Research Fund (FNR) PEARL program, grant agreement 16544475. This work has been partially funded by the Spanish government (PID2023-147592OB-I00, project SE4GenAI ).
\end{acks}

\bibliographystyle{ACM-Reference-Format}
\bibliography{bibliography.bib}

%\appendix
%\section{Appendix1}

\end{document}